# Navigating AI in Social Work and Beyond
# A Multidisciplinary Review

Dalziel, Matt Victor*

Schaffer, Krystal

Martin, Neil

*School of Psychology and Well-being, University of Southern Queensland, Ipswich, Australia.*

*Provide full correspondence details here, including e-mail for the corresponding author.*

vicdalziel@gmail.com

**Abstract**

This review began with the modest goal of drafting a brief commentary on how the social work profession engages with and is impacted by artificial intelligence (AI). However, it quickly became apparent that a deeper exploration was required to adequately capture the profound influence of AI, one of the most transformative and debated innovations in modern history. As a result, this review evolved into an interdisciplinary endeavour, gathering seminal texts, critical articles, and influential voices from across industries and academia. This review aims to provide a comprehensive yet accessible overview, situating AI within broader societal and academic conversations as 2025 dawns. We explore perspectives from leading tech entrepreneurs, cultural icons, CEOs, and politicians alongside the pioneering contributions of AI engineers, innovators, and academics from fields as diverse as mathematics, sociology, philosophy, economics, and more. This review also briefly analyses AI's real-world impacts, ethical challenges, and implications for social work. It presents a vision for AI-facilitated simulations that could transform social work education through *Advanced Personalised Simulation Training (APST).* This tool uses AI to tailor high-fidelity simulations to individual student needs, providing real-time feedback and preparing them for the complexities of their future practice environments. We maintain a critical tone throughout, balancing our awe of AI's remarkable advancements with necessary caution. As AI continues to permeate every professional realm, understanding its subtleties, challenges, and opportunities becomes essential. Those who fully grasp the intricacies of this technology will be best positioned to navigate the impending AI Era.

Since its emergence in 1955, artificial intelligence (AI) has been one of human history's most debated and contentious technological breakthroughs. AI differs from previous technological innovations and disruptions by its potential to replicate, augment, and exponentially improve capacities once exclusive to humans, such as perception, cognition and language. Perspectives on AI fluctuate wildly from viewing it as an existential threat to considering it humanity's saviour. These extreme positions are not limited to sensationalist media, political figures, science fiction, or celebrity academics; AI's leading pioneers also express them. For example, Sam Altman, co-founder and CEO of OpenAI, famously remarked in 2015 that "AI will probably...lead to the end of the world, but in the meantime, there'll be great companies" (Galef, 2015, para. 2). From a more philosophical perspective, Yuval Noah Harari cautions, "If we are not careful, we might be trapped behind a curtain of illusions, which we could not tear away – or even realise is there" (Harari, 2023, para. 16). Sam and Yuval's quotes beautifully encapsulate the tension between society's fear of AI and the eagerness of tech giants and governments to harness its commercial and military potential. This commentary will explore these opposing views, highlight professional opportunities within the terrain between them, examine social work's role in navigating this unprecedented space, and note potential impacts on the profession.

Concerns over AI's rapid development have gained significant traction in tech and public domains. In March 2023, high-profile figures, including Elon Musk and Apple Co-Founder Steve Wozniak, co-authored the widely circulated 'Pause Giant AI Experiments: An Open Letter' (Future of Life Institute, 2023). The letter advocating for a six-month moratorium in AI research gathered over 33,700 signatures, many from within the tech industry. It highlighted mounting fears about the safety of unchecked AI development and its potential societal impact. These anxieties were further substantiated by the 'The Potentially Large

Effects of Artificial Intelligence on Economic Growth' report, published by Goldman Sachs, which projects that up to 300 million jobs globally could be lost due to AI-driven automation (Briggs & Kodnani, 2023). Despite this looming disruption, many within social work have yet to engage in an evidence-based review and discussion on how AI might reshape the profession or how social work can address the fallout of an unprecedented labour market shift (Hodgson et al., 2022; Magruder, 2023). Rather than adopting a *'wait and see'* approach, it is essential for social work to actively shape AI's influence in this evolving context.

Job losses due to AI are already impacting various industries. In February 2024, Hollywood actor and producer Tyler Perry cancelled an $800 million expansion of his Atlanta movie studio after viewing OpenAI's text-to-video software, Sora, in action (Kilkenny, 2024). Stunned by Sora's production of cinematic-quality video, Perry voiced concern over its potential consequences on the entertainment industry, stating:

> I think it's going to be a major game-changer... I am very, very concerned that in the near future, a lot of jobs are going to be lost...There's got to be some sort of regulations in order to protect us. If not, I just don't see how we survive. (Kilkenny, 2024, para. 3-12).

While Perry's concerns about AI's long-term effects on film production are well-founded, his decision to cancel the studio expansion immediately led to the loss of thousands of jobs in construction and related sectors. This exemplifies how AI's impact is not just theoretical but already has tangible repercussions across all sectors of the economy.

Although some may dismiss the actions of Perry and Musk as mere jostling for clicks, subscribers, investor funds, or market dominance, they underscore the disruptive societal impacts of AI. Perry's remarks draw attention to a unique concern: unlike prior technological revolutions that cyclically devastated the working class, a phenomenon Schumpeter (2013) referred to as *Kondratiev Waves*, AI now threatens to disrupt middle-class professions. This shift aligns with Brynjolfsson & McAfee's (2014) concept of the 'Second Machine Age,' driven by AI and robotics, where the physical, digital, and biological realms converge. What distinguishes this 'age' from previous ones is the unprecedented speed at which it is unfolding and the likelihood that AI will make vast segments of human labour obsolete (Brekelmans & Petropoulos, 2020; Frey & Osbourne, 2017). Combining AI with robotics, particularly AI's ability to build, repair and even program itself autonomously, significantly reduces the potential for new jobs and industry creation. This profound shift in human-labour relations has serious ethical, practical, and epistemological implications, particularly for professions such as social work.

Although the picture for many professions may seem bleak, there is broad agreement that roles requiring high emotional intelligence and direct human interaction, such as social work, will survive this redundancy, at least in the short term (Huang & Rust, 2018). Social work has long recognised that human lives are inherently messy, relationships are complex, and life is unpredictable (Healy, 2022). Despite advancements in AI, it remains incapable of replicating many essential aspects of the human condition, including emotional intelligence, empathy, spirituality, kinship bonds, and discrimination and power dynamics in human affairs. (Brynjolfsson & McAfee, 2014; Fernando & Ranasinghe, 2023; Goleman, 1996; Lewis et al., 2018). However, AI's automation of administrative tasks, such as scheduling appointments, tracking case progress reports, and updating case notes and treatment plans (Dey, 2023), could significantly reduce the burden of routine duties, freeing up more time for client-centred, human-focused work (Werder, 2023). Social work's emphasis on human connection ensures that it will remain indispensable as society confronts the challenges posed by AI.

---

AI is poised to compel humbleness within humanity (Bostrom, 2014). For the first time, humans may no longer be the most intelligent entity in the room, forcing society to confront deep existential and philosophical questions, such as "What defines life?". Historically, religious doctrines have shaped societal norms and beliefs; today, this role is increasingly assumed by global technology corporations, whose influence continues to expand (Zuboff, 2019). The potential development of Artificial General Intelligence (AGI) that includes well beyond human levels of cognition in most domains of life represents one of the most significant technological disruptors in history (Mokyr, 1992). Whether this disruption will lead to overwhelmingly positive or negative outcomes, or something in between remains highly contested. As OpenAI (2015, para. 6) aptly stated in its launch document: "It's hard to fathom how much human-level AI could benefit society, and it's equally hard to imagine how much it could damage society if built or used incorrectly". Like electricity, AI has the potential to infiltrate nearly every aspect of human life, transforming societies in ways that are currently unimaginable (Gruetzemacher & Whittlestone, 2019; Šucha & Gammel, 2021).

Despite its growing influence, AI lacks a universally accepted definition. Broadly, it is an advanced technology designed to simulate human intelligence, solve problems, and augment human activities (Stryker & Kavlakoglu, 2024). Like foundational innovations such as the Internet or the combustion engine, AI is often regarded as a general-purpose technology with applications across various domains (Suleyman & Bhaskar, 2023). An unprecedented aspect of AI is its capacity for learning, adaptation, and content creation, with the eventual goal of achieving consciousness. At its core, AI systems consist of four main components: (a) architecture, (b) inputs, (c) algorithms for training on both existing and new data, and (d) outputs (Feuerriegel et al., 2024). Recent advancements, particularly the development of large language models (LLMs) like ChatGPT, have significantly expanded AI's capabilities, enabling it to process and respond to natural language with remarkable sophistication. LLMs are the foundation for next-generation AI systems, including tools like Sora (Naveed et al., 2023).

A long history of gradual and incremental changes has driven AI's evolution. However, a series of recent breakthroughs propelled AI into the advanced deep-learning systems we recognise today. The advent of convolutional neural networks (CNNs) (Krizhevsky et al., 2012) is credited with kickstarting modern deep learning, a foundation that Goodfellow, et al. (2016) further built upon. The introduction of the Transformer architecture (Vaswani et al., 2017), along with BERT for pre-training deep bidirectional transformers in natural language understanding (Devlin et al., 2018), and DeepMind's use of deep reinforcement learning to achieve human-level performance in Atari games (Mnih et al., 2013), represent pivotal moments. These five seminal developments mark critical advances in neural networks, deep learning, and language models, widely recognised as a cornerstone in shaping contemporary AI systems.

Like most technologies, AI is neither inherently good nor evil; instead, it reflects the ideas, ideologies, and biases of those who develop and control it. Consequently, AI has faced numerous legitimate concerns, including issues related to privacy (Reamer, 2013; Roche et al., 2023), bias (Bussey, 2022; Sutaria, 2022), elitism (McQuillan, 2022; Shah, 2024), sexism (Hong et al., 2020; Manasi et al., 2022) and racism (Hong & Williams, 2019; Mathiyazhagan et al., 2022). Additionally, the lack of diversity (Coyle, 2019; Fosch-Villaronga & Poulsen, 2022) and cultural awareness (Lewis, 2023; Munn, 2024) in AI systems raises significant concerns. The environmental impact of AI (Coleman, 2023; Van Wynsberghe, 2021; Wu et al., 2022), implications for human rights (Gaumond & Régis, 2023; Reamer, 2013), and potential threats to democracy (Barnhizer & Barnhizer, 2019) have also come under scrutiny.

Furthermore, AI has also been characterised as authoritarian (McQuillan, 2022), colonial (Adams, 2021; Lewis, 2023; Mohamed et al., 2020), and extractivist (Ricaurte, 2019; 2022), while also facilitating digital pathways to incarceration (Patton et al., 2021). These concerns are exacerbated by the global absence of robust regulatory and ethical frameworks governing AI (Roche et al., 2023; Smuha, 2021). The lax, 'let it rip' approach poses significant risks, particularly for society's most vulnerable populations. These oppressive outcomes of AI conflict with the human rights that social work aims to advocate and protect (Australian Association of Social Workers, 2023), highlighting the profession's critical role in shaping AI's future developments. Without regulatory oversight, AI companies, many of them large multinational corporations, are left to 'self-monitor' their practices regarding equality, transparency, accountability, and data management, which is far from ideal. These corporations often prioritise innovation and profit over the ethical implications of their technologies. Lacking external oversight, legitimate questions arise regarding their accountability, particularly as AI systems become more autonomous.

The limitations of AI discussed thus far assume that humans remain in control, which may only be the case for a short time. Jürgen Schmidhuber, a pioneer in AI and the Scientific Director of the Dalle Molle Institute for Artificial Intelligence Research, expressed this concern in Tonje Schei's (2019) documentary *iHuman: AI and Humanity*. Schmidhuber envisions a future where AI surpasses human intelligence, openly admitting his long-standing ambition: "When I was a boy, I thought, how can I maximise my impact? I have to build something that learns to become smarter than myself". This candid reflection encapsulates a broader ambition within the AI industry: from its inception, AI was designed to evolve beyond human comprehension, potentially escaping human control altogether.

This shift in control raises profound questions, most notably posed by AI researcher Roman Yampolskiy (2024, p. 2), who asks: "How can humanity remain safely in control while benefitting from a superior form of intelligence?". This question's ethical and existential risks remain at the forefront of AI discourse, particularly given the unprecedented pace of technological advances. Yampolskiy (2024) formulated three major obstacles to maintaining control over such systems:

> **1. Unexplainability:** The inability to fully explain or understand all of AI's decisions with 100% accuracy and certainty.
> **2. Unpredictability:** The inability to predict all the pathways and actions an intelligent AI system will take to achieve its goal, even when the goal is known to the observer.
> **3. Non-Verifiability:** The inability to verify, with absolute confidence, the correctness of the AI system's actions, even with probabilistic assessments.

Yampolskiy's challenges imply that AI systems already behave in ways their creators do not entirely understand. It may surprise many that even AI developers and researchers do not fully grasp machine learning algorithms, with some deep learning algorithms remaining well beyond comprehension (Dickson, 2023; Xiang, 2022). For the field of social work, this highlights the need for vigilance, recognising AI as a tool designed by humans but increasingly autonomous and unpredictable. Social workers must not hold AI in awe; instead, they should lean into the profession's expertise in critically reflective practice, and examine its motives, biases, and reasoning. AI should be integrated thoughtfully into practice while remaining aware of potential unintended consequences for professionals and clients.

---

Although AI may seem intangible, existing in a largely invisible virtual space often referred to as Black or Glass Boxes (Rai, 2020; Ribeiro et al., 2016), the physical devices required for its development, maintenance, and use are highly resource-intensive. For example, the reliance on lithium-ion batteries underscores AI's dependence on finite natural resources. The human toll in sourcing these finite resources, particularly cobalt, is a frequently overlooked aspect of AI's rapid advancement. More than 2,000 'creuseurs,' or unregulated cobalt miners, perish annually in the Democratic Republic of Congo, with many victims being children (Amnesty International, 2016, Calvão et al., 2021). Moreover, AI's thirst for electricity is insatiable. Global electricity consumption has surged due to the rapid growth of AI-related industries, which now account for approximately 2% of worldwide electricity use, a figure expected to double within the next two years (Dyck et al., 2024).

Developing advanced AI models and constructing hyper-scale data centres necessary to support these systems have pushed electricity consumption beyond typical industry projections, straining power grids worldwide (Green et al., 2024). Over the past four to five years, companies heavily invested in AI, like Microsoft and Google, have experienced sharp increases in carbon emissions, surging by 29% and 48%, respectively (Kerr, 2024; Milmo, 2024). This trend forced Google to concede that the future environmental impact of AI "...is complex and difficult to predict" (Braue, 2024, para. 5). Despite these concerns, pursuing dominance in the highly lucrative AI market continues to take precedence over addressing environmental issues and protecting human rights.

This short-term drive for control is deeply rooted in anthropocentric philosophies, particularly those emerging from Judeo-Christian thought and Enlightenment ideals (Mignolo, 2011; Tuhiwai Smith, 1999). These traditions emphasise human dominance over nature and the pursuit of objective knowledge by reducing complex systems into their constituent parts. Similarly, AI developers often operate under the assumption that the world is 'knowable' and can be computationally simulated (Griffin et al., 2024). Such perspectives help explain why AI developers and programmers believe they have a 'right' to harvest all publicly available online content to train their models, often arguing that training AI is impossible without incorporating copyrighted content (Gray, 2024; Kazaz, 2024). This content, especially in explainable AI systems (XAI), is also predominantly rooted in Western cultural frameworks, rendering the outputs, programs, and communications biased toward Western populations (Peters & Carman, 2024). Thus, AI continues to emulate and reproduce its developers' narrow ideologies, practices, and concerns (Adams, 2021; Mohamed et al., 2020).

In practice, AI systems are primarily based on discrete event simulations and operate on principles like 'common things happen commonly.' They learn and adapt like Bayes' theorem, updating predictions as new data or evidence becomes available (Bayes, 1763). However, AI's capacity to create order from chaos, much like the stabilising effects described in Lyapunov's equations, remains limited (Lyapunov, 1892/1992). While AI can handle minor disruptions and adapt within structured environments, it struggles to process the messiness of the natural world (Rainie et al., 2021). Processes such as building trust, fostering imagination and curiosity, and experiencing transcendence remain uniquely human. As a profession centred on the human condition in all its complexity, social work is well-positioned to thrive in an AI-driven future.

Despite the profound concerns surrounding AI, it is crucial to acknowledge its promising opportunities, particularly in social work education, advocacy, and client outcomes. AI is already reshaping the profession, offering tools that streamline communication and case management processes (Kumar & Singh, 2023). For example, AI's ability to swiftly and accurately process risk assessments can be invaluable in assisting individuals during crises. Moreover, it holds the potential for identifying and addressing systemic biases within social service delivery, which is a critical issue in creating an equitable practice paradigm (Reamer, 2023). AI can improve communication and safety by identifying and predicting harmful online content, as demonstrated by Patten et al. (2021), who highlight AI's capacity to process nuanced and complex language and cultural contexts—a valuable tool in multicultural societies such as Australia. While AI's ability to analyse large qualitative and quantitative datasets (Siiman et al., 2023), automate routine administrative tasks (Akram, 2024), and predict social worker burnout (Reamer, 2023) could significantly enhance the efficiency and effectiveness of social work practices. Although these technologies are still in their infancy, initial indications suggest that AI could facilitate a more holistic and client-focused approach to social work by alleviating the administrative burden and enhancing practitioner learning.

---

The rapid pace of AI's evolution in education has profoundly impacted teaching institutions and their governing bodies worldwide. AI is contributing to the decline of traditional didactic teaching methods while simultaneously creating opportunities to enhance social work education. Responses to these developments within academia and professional bodies remain divided: some embrace new technologies and tools with an eager openness, while others remain defensive, perceiving such innovations as threats to established practices, particularly in traditional assessment methods. Exploring how these advancements can benefit educators and students is crucial as the field progresses. One promising area is using immersive and interactive simulations, where AI can enhance students' learning experiences through virtual reality (VR) simulations.

Social work environments are often complex and emotionally demanding, contributing to burnout, especially among newly qualified practitioners (Kinman & Grant, 2010; Rose & Palattiyil, 2020). This underscores the urgent need for training programs that introduce students to the realities of professional practice early in their degree programs. While social work can be deeply fulfilling, it may not be the ideal career for everyone. Unfortunately, many students do not realise this until their first field placements, typically during their third year of study. AI simulations offer early opportunities for students to experience social work practice in a controlled environment, enabling them to make more informed decisions about their career paths earlier. From a social justice perspective, this early exposure can prevent students from facing emotional, physical, or psychological trauma, as well as the significant financial costs of pursuing a degree that may not align with their aspirations.

Integrating AI and simulations into social work curricula provides a unique opportunity to address these challenges, particularly in teaching complex topics. These immersive tools promote self-awareness, empathy, reflection, and perspective-taking while enabling students to practice skills such as observation (Jiang & Ahmadpour, 2021). Research demonstrates that simulation-based training can enhance empathy (Pecukonis et al., 2016), improve diagnostic accuracy and clinical interviewing skills (Washburn et al., 2016), and cultivate more positive attitudes toward clients (Lanzieri et al., 2023). Though still an emerging technology, AI-integrated simulations have proven valuable in preparing students for sensitive topics such as child protection and domestic and family violence (Agillias et al., 2021; Jefferies et al., 2023; Schaffer et al., 2024). For example, Touro University in New York has integrated AI-supported clinical training into its curriculum through decision-making trees and VR. This technology is used in simulations involving substance abuse and suicide assessment, allowing students to navigate challenging conversations in a structured, supportive environment (Lanzieri et al., 2021; Mowreader, 2024). AI programs embedded within coursework personalise scenarios, offer real-time feedback, and evaluate performance. Student feedback has been positive, with many appreciating the immediacy of feedback and reported feeling increased confidence in their skills as they prepare to work with real clients.

AI programs like Sora represent a significant milestone in developing simulation learning tools by generating comprehensive and realistic simulations without actors or physical sets. These programs eliminate many logistical challenges and enable large-scale, personalised scenario adjustments to accommodate individual student needs. This advancement holds great potential for cost reduction and reshaping social work education by improving workload management for facilitators, factors often cited as limitations in current simulation designs in higher education (Kourgiantakis et al., 2020; Sewell et al., 2023). AI's capability to identify where a student is struggling and adapt high-fidelity simulation scenarios in real-time to address specific learning needs is what we refer to as *Advanced Personalised Simulation Training (APST).* This approach represents a transformative pedagogical advance.

However, implementing APST raises ethical concerns regarding student privacy and algorithmic surveillance. Current systems require data sharing with the AI platform and with the companies that develop and own the software, raising important questions about data management, storage, and confidentiality. Many AI companies are commercially driven and operate under self-regulation, intensifying concerns about handling confidential student information. Therefore, it is crucial for social work educators and universities to co-design and develop these programs in-house, ensuring that sensitive data is securely stored on-site. Additional challenges include technology acceptance among students and instructors and disparities in access to these tools, especially for students from lower socioeconomic backgrounds or those in regional, rural, or remote areas (Dalziel, 2019). While these challenges are significant, they can be mitigated with targeted interventions.

---

Recently, concerns have emerged around the overall ethical direction of AI development. The Asilomar AI Principles stress the responsibility of developers to pursue "...beneficial intelligence" rather than the risk of "...undirected intelligence" (Future of Life, 2017, para. 2). Yet, this idealistic vision contrasts starkly with more recent realities. Dr Ben Goertzel, Chief Computer Scientist at Hanson Robotics, has observed that AI's developmental focus has increasingly shifted toward militarisation, surveillance, and social manipulation, referring to its usage for "spying, brainwashing, or killing" (Goertzel, 2018, para. 8). This shift reflects how AI, once open and accessible, has become tightly controlled by a small group of elite tech engineers, their billionaire financial backers, and governments (Widder et al., 2023). As Foucault (1980) reminds us, power is omnipresent, manifesting through discourses and the control of knowledge. Today, this power struggle is unfolding in the digital sphere, where global elites seek to monopolise AI's potential for their own interests.

Social work professional organisations, NGOs, advocacy groups, independent media, and academic institutions must critically assess AI's societal impact in a balanced manner, avoiding sensationalism. Together, these groups can be crucial in advocating for guidelines and regulations that ensure ethical AI practices. The governance of AI must prioritise serving the common good while preventing its misuse, including weaponisation. Effective oversight requires transparency, accountability, fairness, and a concerted effort to mitigate bias and avoid homogenising AI development and deployment. Achieving these goals will depend on holding companies and governments accountable through rigorous research, public awareness campaigns, policy recommendations, and, most crucially, establishing internationally binding agreements.

It is chilling that no universally binding international agreement on ethical AI practices exists at this late stage in AI development. While frameworks such as UNESCO's *Recommendation on the Ethics of Artificial Intelligence* (2022) and the G20's *AI Principles* (Organisation for Economic Co-operation and Development, 2019) provide guidelines and proposed regulations, these remain voluntary. The European Union has made some progress towards regulation, beginning with the *Coordinated Plan on AI* (European Commission, 2021). This plan was the foundation for the EU's AI Act *[Regulation (EU) 2024/1689]* (EUR-Lex, 2024). This regulation establishes a comprehensive legal framework for AI governance, though many provisions will not be enforced until August 2026. The Act also includes significant exceptions for military applications, government bodies, public authorities, international organisations, non-EU countries, scientific research, and personal use (KPMG, 2024). These exceptions may limit the Act's effectiveness, rendering it more symbolic than impactful. Furthermore, excluding major AI research nations, such as China and Russia, exacerbates the challenge of global AI governance. The need for enforceable international standards and regulations grows more pressing as AI evolves.

This is especially true as AI gains consciousness, self-awareness and emotional capacity. Neuromorphic computing, exemplified by projects like Neuralink's AI-human integrations, creates life-dependent connections between technology and essential human functions (Chaudhary et al., 2017; Nicolelis, 2011). These developments necessitate a reassessment of the implications of private and government ownership of such technologies (Eubanks, 2018; Zuboff, 2019), including the risks of slavery, exploitation and life-sustaining technological dependence (Bostrom, 2014; Bryson, 2020). The potential for AI to exert control over human lives or to be used as a tool of coercion and manipulation should be considered (Cohen, 2019; O'Neil, 2016). Ethical questions regarding AI rights and personhood also need to be addressed, which further complicates the landscape (Bryson, 2018; Gunkel, 2024). Often dismissed as premature or exaggerated, some of these capabilities already exist or will by 2030, making it crucial to begin crafting robust regulatory frameworks now (Baum, 2017; Dubey, 2024; Naudé & Dimitri, 2020; Ziegler, 2024). While confronting and potentially overwhelming, these developments also open new spaces for social work practice and study, particularly in addressing AI's ethical, psychological, and societal impacts.

Though tackling these challenges may seem intimidating, historical precedent exists for successfully confronting global issues. *The Treaty on the Non-Proliferation of Nuclear Weapons* (NPT) (United Nations, 1968) and the *Montreal Protocol on Substances that Deplete the Ozone Layer* (United Nations, 1987) demonstrate that international agreements on critical issues are achievable. However, such progress requires coordinated efforts between the public and private sectors and interdisciplinary collaboration. While individuals, social workers, and academics play essential roles in addressing these challenges, meaningful progress on AI regulation will ultimately depend on leadership from national and international social work bodies. Social workers can engage with AI research fields, such as 'AI Safety and Security' to ensure that ethical considerations are integrated into AI systems from their inception (Yampolskiy, 2018). By collaborating with specialists or developing expertise in AI, social workers can embed social justice principles into the core of AI development through meaningful co-design processes. Beyond this, social workers can serve as ethicists, advocating for oversight, monitoring, and rigorous auditing of AI systems while lobbying for compensation mechanisms to address the harm caused by AI. Crucially, social workers can ensure the voices and contributions of marginalised populations, including First Nations Peoples, are included in AI developments.

Building on the need for greater inclusion in AI development, First Nations Peoples have, in many respects, already taken steps that surpass the current efforts of the social work profession. Recognising the risks of exclusion from the AI narrative, First Nations leaders convened in Hawai'i in 2020 to draft an Indigenous framework for inclusive AI design. The resulting *Guidelines for Indigenous-Centred AI Design,* grounded in traditional knowledge, are a powerful example of AI advocacy (Lewis, 2020). These guidelines emphasise that homogenisation leads to cultural loss (Lewis, 2020) and propose that humanity must learn to form "kin with the machines" (Lewis et al., 2018, p. 1). This perspective calls for a radical departure from dominant paradigms, recognising that humans are part of interconnected systems that include animals, rivers, rocks, wind (Abdilla, 2019), and even machines. The guidelines draw upon the Lakota Seven Generations principle, urging us to consider how our actions today will shape AI for the next seven generations (Kite et al., 2020). Social work organisations can learn from these custodianship principles when developing their ethical frameworks for AI. Co-designing AI systems with marginalised communities and integrating Indigenous perspectives could foster a more inclusive and ethical approach to AI development. Social workers do not need to write code to influence AI; engaging in reflective dialogue and collaboration is sufficient to begin shaping a more just and inclusive future. Ultimately, social workers can help shape this vision or risk having it shaped for them.

---

This commentary began by discussing Elon Musk and his associates' open letter, which highlighted concerns shared by many within and outside the tech industry regarding AI's rapid development. However, it's essential to understand this letter in context; it was not an argument against AI technology but a warning about the dangers of unregulated research and development. While sounding the alarm on AI's more unethical aspects, Musk, Perry, and Wozniak have simultaneously been pioneering and leveraging the latest AI technologies. Since the letter's publication, Musk has launched Grok, a sarcastic AI chatbot backed by $6 billion in funding, to compete directly with the global juggernaut ChatGPT. Wozniak's company, Apple, introduced Apple Intelligence (AI), while Tyler Perry incorporated AI technologies in two of his most recent films. Thus, despite their warnings and proclamations, these global powerbrokers are deeply embedded in AI research and development. Social work practitioners and professional bodies worldwide must adopt this critically reflective and informed stance, proceeding cautiously from an informed position within AI's control centres. It is vital that our discipline does not shy away from AI, but rather acknowledges and engages with it in an authentic and values-based manner.

As this review has aimed to illustrate, AI is at an inflection point, with its potential to disrupt social work and society at large still unfolding. However, the future of AI is not yet set in stone; it remains malleable. Reconciling diverse perspectives while grappling with a rapidly evolving and complex technology can easily overwhelm and confuse. As debates continue to polarise AI as either an existential threat or a transformative saviour, it is crucial to engage in rational, evidence-based discussions to unlock its potential for enriching learning experiences and preparing future social workers for the complexities of a technologically integrated world. Ideally, social work academics should be involved in AI's design from its inception, not merely consumers of the final product. Researchers, social workers, and citizens are responsible for shaping this remarkable innovation to reflect the rich diversity of the human experience and the complexity and unpredictability of the human condition. Whether the future of AI is driven by "killing, spying and brainwashing" or dedicated to education, human rights and social justice is ultimately up to us all. It's time for social workers, sociologists, and academics to claim a seat at the AI table.

- 22 -

Stryker, C., & Kavlakoglu. E. (2024, August 16). *What is AI?* IBM. https://www.ibm.com/topics/artificial-intelligence

Šucha, V., & Gammel, J. P. (2021). *Humans and Societies in the Age of Artificial Intelligence.* European Commission: Directorate-General for Education, Youth, Sport and Culture. https://ai4si.gzs.si/uploads/AlBzFQ1y/HumansandSocietiesintheAgeofAI.pdf

Suleyman, M., & Bhaskar, M. (2023). *The coming wave: Technology, power, and the twenty-first century's greatest dilemma.* United States: Penguin Random House.

Sutaria, N. (2022). Bias and ethical concerns in machine learning. *ISACA J., 4*, 1-4.

Tuhiwai Smith, L. (1999). *Decolonizing Methodologies.* London: Zed Books.

UNESCO. (2022). *Recommendation on the Ethics of Artificial Intelligence: Document Code: SHS/BIO/PI/2021/1.* The United Nations Educational, Scientific and Cultural Organization. https://unesdoc.unesco.org/ark:/48223/pf0000381137

United Nations. (1968, June 12). *Treaty on the Non-Proliferation of Nuclear Weapons*. Office of Legal Affairs: The United Nations. https://legal.un.org/avl/ha/tnpt/tnpt.html

United Nations. (1987, September 16). *A Montreal Protocol on Substances that Deplete the Ozone Layer.* United Nations Treaty Collection. https://treaties.un.org/pages/ViewDetails.aspx?src=TREATY&mtdsg_no=XXVII-2-a&chapter=27&clang=_en

Van Wynsberghe, A. (2021). Sustainable AI: AI for sustainability and the sustainability of AI. *AI and Ethics, 1*(3), 213-218.

Vaswani, A., Shazeer, N., Parmar, N., Uszkoreit, J., Jones, L., Gomez, A. N., Kaiser, L., & Polosukhin, I. (2017). Attention is all you need. *Advances in Neural Information Processing Systems, 30*.

Washburn, M., Bordnick, P., & Rizzo, A. S. (2016). A pilot feasibility study of virtual patient simulation to enhance social work students' brief mental health assessment skills. *Social work in health care*, *55*(9), 675-693.

Werder, C. (2023, September 28). *The Importance of Emotional Intelligence in the Age of AI.* ei Design: Powered by MPS. https://www.eidesign.net/emotional-intelligence-in-the-ai-age/

Widder, D. G., West, S., & Whittaker, M. (2023, August 18). *Open (for business): Big tech, concentrated power, and the political economy of open AI.* SSRN. https://ssrn.com/abstract=4543807

Wu, C. J., Raghavendra, R., Gupta, U., Acun, B., Ardalani, N., Maeng, K., ... & Hazelwood, K. (2022). Sustainable ai: Environmental implications, challenges and opportunities. *Proceedings of Machine Learning and Systems, 4,* 795-813.
- 29 -

Xiang, C. (2022, November 1). *Scientists Increasingly Can't Explain How AI Works.* VICE Magazine. https://www.vice.com/en/article/scientists-increasingly-cant-explain-how-ai-works/

Yampolskiy, R. V. (Ed.). (2018). *Artificial intelligence safety and security*. Florida: CRC Press.

Yampolskiy, R. V. (2024). *AI: Unexplainable, Unpredictable, Uncontrollable.* Florida: CRC Press.

Ziegler, B. (2024, September 21). *It's the Year 2030. What Will Artificial Intelligence Look Like?* The Wall Street Journal. https://www.wsj.com/tech/ai/future-of-ai-2030-experts-654fcbfe#

Zuboff, S. (2019). *The Age of Surveillance Capitalism.* United Kingdom: Profile.

- 30 -